\documentclass[10pt,twocolumn,letterpaper]{article}

\usepackage{iccv}
\usepackage{times}
\usepackage{epsfig}
\usepackage{graphicx}
\usepackage{amsmath}
\usepackage{amssymb}
\usepackage[accsupp]{axessibility}

\usepackage{hyperref}
\hypersetup{breaklinks=true,colorlinks,bookmarks=false}
\usepackage[capitalize]{cleveref}
\usepackage{adjustbox}
\usepackage{booktabs}
\usepackage{multirow}

\iccvfinalcopy 

\def\iccvPaperID{30} 

\ificcvfinal\fi

\begin{document}

\title{Robust Interactive Semantic Segmentation of Pathology Images with Minimal User Input}

\author{Mostafa Jahanifar \and
Neda Zamani Tajeddin \and Navid Alemi Koohbanani \and Nasir Rajpoot\\
Tissue Image Analytics Centre, Department of Computer Science, University of Warwick, UK\\
{\tt\small mostafa.jahanifar@warwick.ac.uk}
}

\maketitle
\ificcvfinal\thispagestyle{empty}\fi

\begin{abstract}
   From the simple measurement of tissue attributes in pathology workflow to designing an explainable diagnostic/prognostic AI tool, access to accurate semantic segmentation of tissue regions in histology images is a prerequisite. However,  delineating different tissue regions manually is a laborious, time-consuming and costly task that requires expert knowledge. On the other hand, the state-of-the-art automatic deep learning models for semantic segmentation require lots of annotated training data and there are only a limited number of tissue region annotated images publicly available. To obviate this issue in computational pathology projects and collect large-scale region annotations efficiently, we propose an efficient interactive segmentation network that requires minimum input from the user to accurately annotate different tissue types in the histology image. The user is only required to draw a simple squiggle inside each region of interest so it will be used as the guiding signal for the model. To deal with the complex appearance and amorph geometry of different tissue regions we introduce several automatic and minimalistic guiding signal generation techniques that help the model to become robust against the variation in the user input. By experimenting on a dataset of breast cancer images, we show that not only does our proposed method speed up the interactive annotation process, it can also outperform the existing automatic and interactive region segmentation models.
\end{abstract}

\renewcommand{\thefootnote}{\fnsymbol{footnote}}
\footnotetext[0]{\copyright \space 2021 IEEE. Personal use of this material is permitted. Permission from IEEE must be obtained for all other uses, in any current or future media, including reprinting/republishing this material for advertising or promotional purposes, creating new collective works, for resale or redistribution to servers or lists, or reuse of any copyrighted component of this work in other works.}

\section{Introduction}
\label{sec:intro}

Annotated data in computational pathology (CPath)  plays an important role in developing algorithms for cancer diagnosis \cite{mahmood2021artificial} 
and prognosis \cite{alsubaie2021tumour}. 
Measurement of tissue attributes can also help make the diagnostic process more objective and reproducible. 
Furthermore, annotations in digital pathology can be leveraged to extract hidden information embedded in images which cannot be perceived by the human eye but may be of significant diagnostic value \cite{zamanitajeddin2021sna}. 

In digital pathology with the rise of utilizing digital scanners in hospitals and cancer diagnosis centres, digital images are generated at an unprecedented pace.  However, providing accurate and reliable labels for these large data repositories is not feasible because it requires expert knowledge, is very time consuming and labour-intensive. Pixel-wise (dense) annotation of tissue regions, which is necessary for quantification and increasing the explainability of models \cite{binder2021explainable}, is one of the hardest type of annotations to obtain in digital pathology due to the large size of digital whole slide images (WSIs) that increases the search area for the annotator. Besides, it has been shown that  pixel-wise concordance between tissue type annotations from different human annotators in the same region is not very high  \cite{wahab2021semantic}, which portend the challenging nature of the task.

CPath can facilitate this procedure by identifying and segmenting different tissue elements such as cells \cite{jahanifar2016automatic}, nuclei \cite{koohababni2018nuclei, kumar2019multi, koohbanani2019nuclear}, glands \cite{graham2019mild}, cancer regions \cite{vu2019methods}, etc, for down-stream analysis tasks \cite{zhang2020predicting}. 
However, deep learning (DL) models at the forefront of CPath algorithms require a large amount of annotated data to be trained effectively, otherwise, they cannot generalize well on unseen data \cite{lecun2015deep}. Recently,  human-in-the-loop annotation approaches have been proposed in the literature \cite{budd2021survey} where a DL model generates initial annotations and then the results are reviewed and refined by pathologists. The DL model is then tuned on the refined data and applied to the images to propose a new and presumably better set of annotations \cite{budd2021survey}.
Nonetheless, implementing this approach for tissue region annotation task would not be feasible yet as there is no large publicly available dataset with tissue regions annotated, consequently, the initial DL model may not work well and the need for results refinement will increase. Therefore, the overall human-in-the-loop annotation process might take more time than manual annotation.

One way of facilitating and expediting the process of providing dense annotation is developing platforms and software for interactive segmentation of digital images which demand minimum human effort and time \cite{koohbanani2020nuclick, jahanifar2019nuclick}. 
Utilizing such software can help us to save time
and also generate a large number of annotations for downstream analysis or quantification \cite{gamper2020pannuke}. Benefiting from human input as guiding signals before segmentation, makes interactive models very accurate and robust in segmenting objects in images from new datasets or domains \cite{koohbanani2020nuclick}.

In this paper, we propose an interactive segmentation model, in which the user provided squiggles guide the model toward semantic segmentation of tissue regions. Particularly, our main contributions in this work are listed here:
\begin{itemize}
    \item \textbf{Efficient-UNet}: a light-weight, scaleable, and efficient network architecture that
    benefits from a novel Residual Multi-Scale (RMS) block to better recognize regions with various sizes.
    \vspace{-5pt}
    \item \textbf{Minimalistic guiding signals}: four novel techniques are proposed to allow automatic generation of human-drawn-like guiding signal for training of interactive semantic segmentation model. These methods help the model to become robust against the variations in user input.
    \vspace{-5pt}
    \item State-of-the-art performance on the largest publicly available dataset \cite{amgad2019structured} for semantic tissue region segmentation. 
\end{itemize}

\section{Related work}
\label{sec:related}

\begin{figure*}[!ht]
\begin{center}
\includegraphics[width=0.82\linewidth]{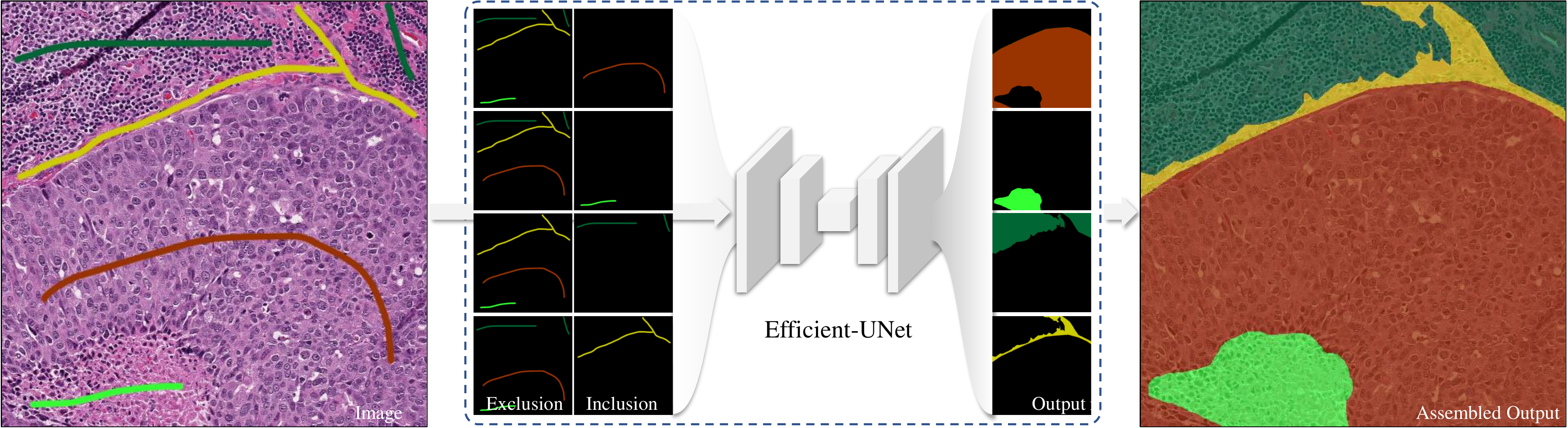}
\end{center}
   \caption{Overview of the proposed interactive segmentation framework. Squiggles in different colours specify different region types, based on which \textit{inclusion} and \textit{exclusion} guiding signals are generated to be fed into the network and generate segmentation map.
   }
\label{fig:overview}
\end{figure*}

In various works interactive object segmentation is formulated as energy minimization on a graph defined over the objects \cite{ramadan2020survey, rother2004grabcut}. However, it has been shown in \cite{koohbanani2020nuclick} that these kinds of methods cannot handle the complexity that might appear in histology primitives (glands or tissue regions) and their performance in comparison to supervised and DL-based methods is relatively poor.  
  
Deep learning models have also been extensively used for interactive segmentation \cite{ramadan2020survey, xu2017deep, maninis2018deep, ling2019fast, castrejon2017annotating} among which DEXTRE \cite{maninis2018deep} is a powerful approach that utilizes extreme points as an auxiliary input to the network. First, the annotator clicks four points on the extreme positions of objects then a heat map (Gaussian map for each point where points are at the centres of Gaussian maps) channel is created from these clicks which are attached to the input and serve as guiding signals. Although DEXTRE performs very well on natural images \cite{maninis2018deep}, this kind of approach does not segment different tissue types in histology images for several reasons. First, there may be several separate regions belonging to the same region type, where selecting extreme points for all of them together is not feasible. Furthermore, tissue regions from different tissue types are usually entangled or nested (see \cref{fig:results} Ground Truth column). Therefore, the extreme points for different regions may overlap with or fall close to (within) each other which may confuse DEXTRE in recognizing different regions.

There are other methods in the literature that require the user to draw a bounding box around the desired object as a guiding signal for the model. Wang \etal \cite{wang2018interactive} applied a deep network on a cropped image based on a bounding box to obtain segmentation. In a correction phase, this approach also takes squiggles from the user to indicate the foreground and background. However, this model is not practical for segmentation of multiple objects (like nuclei) or amorphous objects (like glands or tissue region) in the histology domain. Some methods combine bounding box annotations with Graph Convolutional Network (GCN) to achieve image segmentation \cite{ling2019fast,castrejon2017annotating} where the selected bounding box is cropped from the image and fed to a GCN to predict polygon/spline around the object. Again, similar to extreme point-based signals, methods relying on bounding box signals cannot address the shape complexity, nested objects and entangled regions challenges that are present in histology images. 

To the best of our knowledge, NuClick \cite{koohbanani2020nuclick} is the only interactive segmentation approach for extracting objects in histology images in the literature that deals with these challenges by introducing the use of squiggle based guiding signals.  In the original NuClick \cite{koohbanani2020nuclick}, a random point inside the GT mask and morphological skeleton of the GT mask was used for nucleus and gland segmentation tasks, respectively.

Although the morphological skeleton of the original mask may work well in the case of gland instance segmentation \cite{koohbanani2020nuclick}, for semantic region segmentation task the skeleton based on the original mask may get complicated due to the noise and variations in the boundaries of the original mask. This effect can be seen in \cref{fig:signal}(a), where the morphological skeleton of the original mask is extracted using Lee \etal method \cite{lee1994skeleton} and it has many small branches that are over-fitted to the GT mask. We have addressed this issue in this work by introducing \textit{minimalistic guiding signals}.


\section{The proposed method}
\label{sec:methods}

The pipeline of the proposed interactive semantic segmentation framework is illustrated in \cref{fig:overview}. Our segmentation network follows an encoder-decoder paradigm where two auxiliary maps termed as the ``\textit{guiding signals}'' are concatenated to the input image. One of these maps point to the desired object in the image (\textit{inclusion map}) and another one refers to the other objects that are present in the same field of view (FOV) but are not of interest (\textit{exclusion map}). During the training phase of the network, these guiding signals are automatically generated based on the ground truth mask whereas in the inference phase we expect the user to draw the guiding signals interactively. In either case, we expect a fine segmentation of the desired object, which is marked by the inclusion signal, in the model output. Outputs related to all guiding signals are then assembled to form the final semantic segmentation map.
In the current work, we propose four techniques to make the model more robust against the variation in the guiding signals by modifying the extracted morphological skeleton during the training (\cref{sec:skeleton}) and introduce an efficient network architecture that is fast and accurate in segmentation task (\cref{sec:model}). 

\subsection{Minimalistic guiding signals}
\label{sec:skeleton}

\begin{figure*}[th!]
\begin{center}
\includegraphics[width=0.80\linewidth]{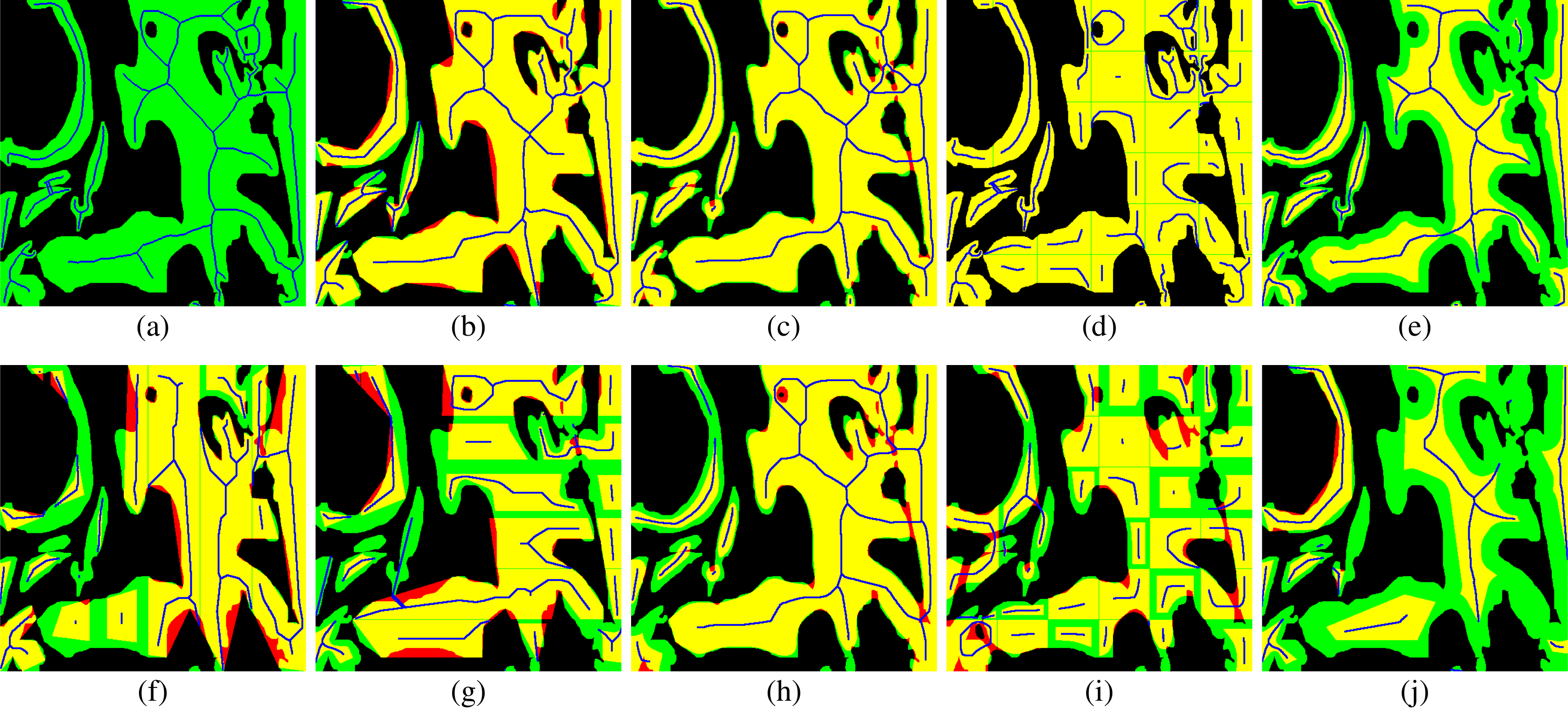}
\end{center}
   \caption{Minimalistic guiding signal generation techniques. (a) shows the original (GT) tissue mask in green and its corresponding skeleton in blue colour. Images (b)-(e) illustrate the effect of mask approximation, mask smoothing, mask partitioning, and distance transformation thresholding on the output guiding signal, respectively. In all these images red and yellow regions indicate the modified mask (used for skeleton extraction) and its intersection with the ground truth, respectively. (f)-(e) are five examples of random combination of mask modification techniques and their resulting skeleton.}
\label{fig:signal}
\end{figure*}

\subsubsection{Mask approximating}
One way to reduce the noise on the mask boundaries is to approximate the mask polygon to a similar but more simplistic polygon. For this goal, we extract the boundary coordinates of the binary mask to form its polygon, then apply Douglas–Peucker algorithm \cite{douglas1973approx} to iteratively decimate the original polygon to a similar polygon with fewer points. Then, we convert the approximated polygon back to a binary mask. Doing so, we filter out small changes (noise or details) on the mask boundary to obtain a much more simple skeleton, as depicted in \cref{fig:signal}(b). To better follow the changes related to mask approximation in \cref{fig:signal}, the original mask, the approximated (modified) mask, and their intersection are showed in green, red, and yellow colours, respectively, while the blue line segments depict the generated guiding signal (skeleton) based on the modified mask.
During the training, we randomly change the approximation accuracy (the maximum distance between the original polygon and its approximation) to obtain a different guiding signal in each epoch.

\subsubsection{Mask smoothing}
The idea of image smoothing prior to the medial axis (skeleton) extraction has been investigated before \cite{fritsch1994medialaxis} as it would result in less noisy skeletons. In this research, we adopt the same idea and apply smoothing filters on the original mask to reduce the noise in the boundaries and then attain a better morphological skeleton. In our implementation, Gaussian and Median filters with variable scales are incorporated. By changing the filter scale (kernel size and sigma) during the training phase, the resulted modified mask (and the generated skeleton) would be different in each epoch. The result of applying a Gaussian filter with kernel size of $25\times25$ and $\sigma=15$ is shown in \cref{fig:signal}(c).

\subsubsection{Mask partitioning}
Due to the approach used for morphological skeleton generation \cite{lee1994skeleton}, the resulted guiding signal usually align with the major axis of the source mask or falls on the centre-line of it (see \cref{fig:signal}(a)-(c)).
In real-world applications, however, the user may draw the guiding signals perpendicular to the main axis or even in segmented lines that do not follow the same orientation. To mimic this behaviour during the guiding signal generation, we propose to partition the mask into smaller parts and calculate the morphological skeleton for each part separately. In our design, each connected component in the mask is treated separately and partitioned adaptively i.e., partitioning direction (horizontal or vertical or both directions) and the number of partitioning lines is decided based on the sizes of that connected component in each axis. 
Partitioned regions do not necessarily follow the same orientation as the main mask, hence, different parts would result in different skeleton representations. You can see the mask partitioning effect on the generated skeletons in \cref{fig:signal}(d) where the biggest object (connected component) in the original mask has been partitioned using a grid pattern (both horizontal and vertical partitioning).  The generated guiding signals via this method are not only minimalistic but also vary significantly which will make the network robust against the vast variation in the user input during the inference.

\subsubsection{Mask distance transform thresholding}
Calculating the distance transformation map of the input mask, randomly thresholding it, and then extracting the guiding signal is the only technique proposed in the  NuClick \cite{koohbanani2020nuclick}. This method proved to be very effective as not only does it smooth the original mask to obtain a more simple and real guiding signal, but also it shortens the length of the generated skeleton which is a valid variation in real-world guiding signals. Having said that, in NuClick \cite{koohbanani2020nuclick}, the distance map was thresholded based on the statistics of all connected components in the mask which was only one gland whereas in tissue region segmentation we have multiple connected components (regions) with various sizes related to the same class and therefore presented in the same mask. If we estimate the threshold based on the statistics of all objects together, that might result in eliminating some objects from the deformed mask and hence no guiding signal would be generated for those objects. Therefore, we proposed to apply distance transformation and thresholding in an adaptive manner to each connected component in the mask, separately. A sample of applying this method to a multi-object mask is depicted in \cref{fig:signal}(e).

\subsubsection{On-the-fly signal generation}
Our implementation of the above-mentioned techniques allows us to incorporate a combination of them for automatic guiding signals generation (both inclusion and exclusion maps) during the training phase. In particular, we apply this ordered sequence of mask approximating, smoothing, partitioning, and distance transform thresholding techniques with probabilities of 0.75, 0.75, 0.5, and 0.5, respectively, and after all, we generate the morphological skeleton \cite{lee1994skeleton} of the modified mask as the guiding signal. The combination of these mask modification techniques will guarantee the generation of unique minimalistic guiding signals in each epoch, as illustrated in \cref{fig:signal}(f)-(j). It is important to note that although a copy of the original mask is modified to generate the guiding signal, the original mask is used for network training as the expected output.

\subsection{Model and loss function}
\label{sec:model}

Inspired by EfficientNet \cite{tan2019efficientnet}, we propose an encoder-decoder network architecture for end-to-end segmentation in which skip connections between the encoder and decoder part have been utilized, similar to UNet \cite{ronneberger2015unet}. We keep the idea of network scalability as suggested in EfficientNet \cite{tan2019efficientnet} and mainly used the same convolutional blocks proposed in that work. However, to make our model robust against variation in object scales, we propose a novel Residual Multi-scale (RMS) block inspired by multi-scale blocks introduced in \cite{jahanifar2018segmentation}. In the following, we explain in detail how these network constructing blocks work, and how we use them to build a scalable segmentation network architecture.

\subsubsection{Network constructing blocks}

\begin{figure*}[!ht]
\begin{center}
\includegraphics[width=0.80\linewidth]{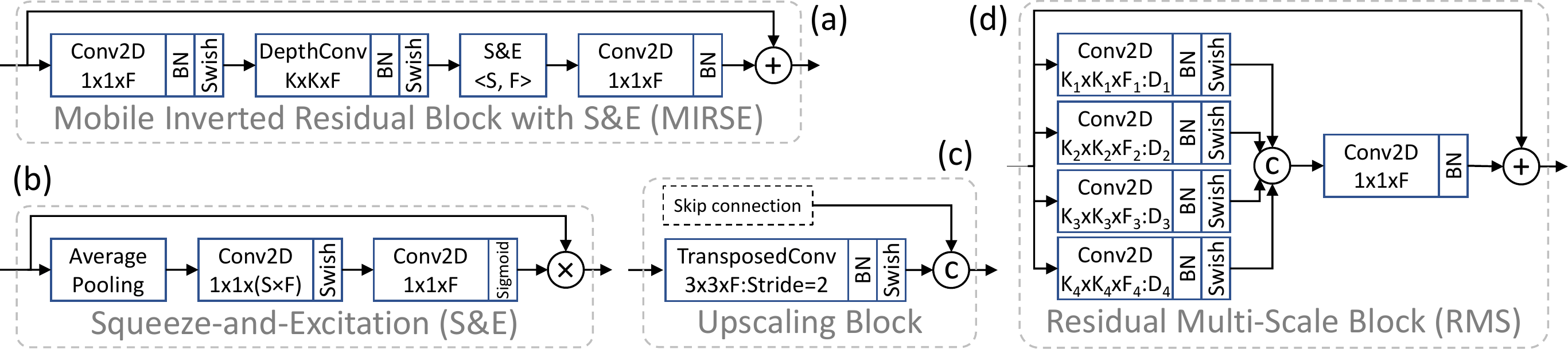}
\end{center}
   \caption{Constructing blocks of the Efficient-UNet segmentation model. 
   }
\label{fig:blocks}
\end{figure*}

\textbf{Mobile Inverted Residual block with Squeeze and Excitation (MIRSE):} These blocks as the main part of the network are directly taken from EfficientNet \cite{tan2019efficientnet}. This block is constructed by embedding a squeeze-and-excitation (S\&E) mechanism \cite{hu2018squeeze} into a residual block from MobileNetsV2 \cite{sandler2018mobilenetv2}. As depicted in \cref{fig:blocks}(a), our implementation of MIRSE blocks requires setting three parameters $K$, $F$, and $S$ that are convolution kernel size, the number of feature maps, and squeeze ratio for the S\&E (\cref{fig:blocks}(b)), respectively.  Note that in each block and after convolution layers, we use Batch Normalization \cite{ioffe2015batch} (BN) and self-gated activation function \cite{ramachandran2017swish} (Swish) layers. Also, note that the squeeze parameter for all MIRSE blocks in our network has been set to $S=0.25$.

\textbf{Upscaling block:} This block consists of a 2D transposed convolution layer with a kernel size of 3 and stride of 2 to increase the resolution of feature maps in each stage of the decoding path. This block is also responsible for concatenating the feature maps from the encoder part of the network and creating skip connections.

\textbf{Residual Multi-Scale block (RMS):} {This block consists of 4 convolution layers (followed by BN and Swish layers) at different scales. Each layer is a \textit{atrous} convolution \cite{chen2017deeplab} with different kernel size ($K$) and dilation rate ($D$). In comparison to MSC block \cite{jahanifar2018segmentation}, the proposed RMS block combines the output feature maps from different scales using a $1\times1$ convolution and incorporates a residual connection from its input to both keep the effect of original input and ease the flow of gradient during backpropagation \cite{he2016resnet}.}

\subsubsection{Efficient-UNet Architecture}
Similar to \cite{tan2019efficientnet}, we first introduce a baseline model architecture and then scale its width (number of channels or feature maps in constructing blocks) and depth (number of block repetition in each stage of network) uniformly using $w$ and $d$ scaling factors, respectively. These factors are calculated using a compound scaling method and are directly adopted from \cite{tan2019efficientnet}. Note that unlike \cite{tan2019efficientnet} we did not scale the network for resolution because the concept of resolution in digital pathology depends on the optical magnification and changing image size or FOV will affect the problem dramatically. 

The architecture of our baseline segmentation model, Efficient-UNet-B0, is described in \cref{tab:net}, in which operations and their configurations in the encoding and decoding paths of the network are outlined. The encoder part contains only MIRSE blocks with different configurations and numbers of repetition, on the other hand, the proposed decoder part encompasses upscaling and RMS blocks as well.
As we have four layers of \textit{atrous} convolution in RMS block, this block is configured using kernel size vector $\mathbf{K}=[K_1, K_2, K_3, K_4]$ and dilation rate $\mathbf{D}=[D_1, D_2, D_3, D_4]$. From \cref{tab:net} note that in stages with *-marked MIRSE block, the first convolution layer would be applied with a stride of 2 to decrease the spatial resolution of feature maps by a factor of 2, similar to max-pooling operation.

During the network scaling, the number of feature maps ($F$ network width) in all blocks (except for the last Conv2D which is responsible for segmentation map generation) and the number of MIRSE block repetitions in each stage ($R$ network depth) will be scaled by $w$ and $d$ factors whereas the size of all kernels and number of Upscale and RMS blocks stay the same.

\begin{table}[!t]
\begin{adjustbox}{max width=\columnwidth}
\begin{tabular}{@{}ll@{}}
\toprule \midrule
\multicolumn{1}{c}{Encoder}                       & \multicolumn{1}{c}{Decoder}                                                                                     \\ \midrule
\multicolumn{1}{l|}{Conv2D ($K$=3, $F$=32, Stride=2)} & \begin{tabular}[c]{@{}l@{}}Upscale ($F$=320)\\ MIRSE ($K$=5, $F$=192, $R$=3)\end{tabular}                               \\ \midrule
\multicolumn{1}{l|}{MIRSE (($K$=3, $F$=16, $R$=1)}      & \begin{tabular}[c]{@{}l@{}}RMS ($\mathbf{K}$={[}3,5,5,7{]}, $\mathbf{D}$={[}3,3,5,7{]}, $F$=192)\\ MIRSE ($K$=5, $F$=112, $R$=3)\end{tabular} \\ \midrule
\multicolumn{1}{l|}{MIRSE ($K$=3, $F$=24, $R$=2)\textsuperscript{*}}       & \begin{tabular}[c]{@{}l@{}}Upscale ($F$=112)\\ MIRSE ($K$=3, $F$=80, $R$=3)\end{tabular}                                \\ \midrule
\multicolumn{1}{l|}{MIRSE ($K$=5, $F$=40, $R$=2)\textsuperscript{*}}       & RMS ($\mathbf{K}$={[}3,3,5,5{]}, $\mathbf{D}$={[}1,3,3,5{]}, $F$=80)                                                                    \\ \midrule
\multicolumn{1}{l|}{MIRSE ($K$=3, $F$=80, $R$=3)\textsuperscript{*}}       & \begin{tabular}[c]{@{}l@{}}Upscale ($F$=80)\\ MIRSE ($K$=5, $F$=40, $R$=2)\end{tabular}                                 \\ \midrule
\multicolumn{1}{l|}{MIRSE ($K$=5, $F$=112, $R$=3)}      & \begin{tabular}[c]{@{}l@{}}Upscale ($F$=40)\\ MIRSE ($K$=3, $F$=24, $R$=2)\end{tabular}                                 \\ \midrule
\multicolumn{1}{l|}{MIRSE ($K$=5, $F$=192, $R$=5)\textsuperscript{*}}      & \begin{tabular}[c]{@{}l@{}}Upscale ($F$=24)\\ MIRSE ($K$=3, $F$=16, $R$=1)\end{tabular}                                 \\ \midrule
\multicolumn{1}{l|}{MIRSE ($K$=3, $F$=320, $R$=1)}                           & Conv2D ($K$=1, $F$=1)                                                                                               \\ \midrule
\bottomrule
\end{tabular}
\end{adjustbox}
\caption{Architecture of the baseline network \textbf{Efficient-UNet-B0}. For each operation, its parameters are outlined in the parenthesis. 
}
\label{tab:net}
\end{table}

\subsubsection{Loss function}
We use a hybrid loss function which consists of soft Dice and binary cross entropy (BCE) parts \cite{jahanifar2020tendon}:
\begin{equation}
    {\mathcal{L} = 1 - \frac{{2\sum\limits_{i,j} {{p_{ij}}{g_{ij}}}  + \varepsilon }}{{\sum\limits_{i,j} {p_{ij}^2}  + \sum\limits_{i,j} {g_{ij}^2}  + \varepsilon }} - \frac{1}{N}\sum\limits_{i,j} {{g_{ij}}\log ({p_{ij}})},}
\label{eq:Loss}
\end{equation}
where ${{p_{ij}}}$ and  ${{g_{ij}}}$ are the values of every (i,j) pixel in the prediction and GT mask, respectively, $\varepsilon=1$ is used to avoid division by zero, and $N$ is the total number of pixels. The Dice part of the loss function is robust against the pixels imbalanced class population whereas BCE penalizes the loss for even small pixel-wise errors \cite{jahanifar2020tendon}.

\section{Experiments and results}
\label{sec:results}

\begin{table*}[ht!]
\begin{adjustbox}{max width=\textwidth}
\begin{tabular}{@{}lllll|lll|lll|lll|lll|lll@{}}
\toprule
\toprule
\multicolumn{1}{c}{\multirow{2}{*}{Model}}          & \multicolumn{1}{c}{\multirow{2}{*}{Guiding Signal}} & \multicolumn{3}{c}{Overall}                                                   & \multicolumn{3}{c}{Tumour}                                                    & \multicolumn{3}{c}{Stroma}                                                    & \multicolumn{3}{c}{Inflamatory}                                               & \multicolumn{3}{c}{Necrosis}                                                  & \multicolumn{3}{c}{Others}                                                    \\ \cmidrule(l){3-20} 
\multicolumn{1}{c}{}                                & \multicolumn{1}{c}{}                                & \multicolumn{1}{c}{DICE} & \multicolumn{1}{c}{Acc.} & \multicolumn{1}{c|}{AUC} & \multicolumn{1}{c}{DICE} & \multicolumn{1}{c}{Acc.} & \multicolumn{1}{c|}{AUC} & \multicolumn{1}{c}{DICE} & \multicolumn{1}{c}{Acc.} & \multicolumn{1}{c|}{AUC} & \multicolumn{1}{c}{DICE} & \multicolumn{1}{c}{Acc.} & \multicolumn{1}{c|}{AUC} & \multicolumn{1}{c}{DICE} & \multicolumn{1}{c}{Acc.} & \multicolumn{1}{c|}{AUC} & \multicolumn{1}{c}{DICE} & \multicolumn{1}{c}{Acc.} & \multicolumn{1}{c}{AUC} \\ \midrule
Amgad \etal \cite{amgad2019structured} & -                                                   & 0.750                    & 0.783                    & 0.898                    & 0.851                    & 0.804                    & 0.941                    & 0.8                      & 0.824                    & 0.881                    & 0.712                    & 0.743                    & 0.917                    & 0.723                    & 0.872                    & 0.864                    & 0.666                    & 0.67                     & 0.885                   \\
UNet \cite{ronneberger2015unet}                                               & -                                                   & 0.733                    & 0.756                    & 0.880                    & 0.833                    & 0.772                    & 0.913                    & 0.788                    & 0.798                    & 0.873                    & 0.701                    & 0.722                    & 0.907                    & 0.713                    & 0.856                    & 0.842                    & 0.629                    & 0.634                    & 0.866                   \\
DeepLab v3 \cite{chen2017deeplab}                                        & -                                                   & 0.760                    & 0.794                    & 0.904                    & 0.862                    & 0.822                    & 0.956                    & 0.812                    & 0.841                    & 0.893                    & 0.725                    & 0.76                     & 0.921                    & 0.729                    & 0.875                    & 0.865                    & 0.671                    & 0.673                    & 0.887                   \\ \midrule
NuClick  \cite{koohbanani2020nuclick}                                           & NuClick                                             & 0.773                    & 0.957                    & 0.968                    & 0.855                    & 0.941                    & 0.972                    & 0.793                    & 0.926                    & 0.957                    & 0.724                    & 0.958                    & 0.960                    & 0.810                    & 0.974                    & 0.983                    & 0.681                    & 0.986                    & 0.967                   \\
NuClick                                             & Minimalistic                                        & 0.835                    & 0.967                    & 0.991                    & 0.895                    & 0.951                    & 0.989                    & 0.839                    & 0.944                    & 0.981                    & 0.824                    & 0.970                    & 0.991                    & 0.851                    & 0.979                    & 0.996                    & 0.768                    & 0.990                    & 0.995                   \\ \midrule
Efficient-Unet-B0 ($w=1, d=1$)                                   & Minimalistic                                        & 0.867                    & 0.978                    & 0.994                    & 0.931                    & 0.973                    & 0.995                    & 0.877                    & 0.961                    & 0.987                    & 0.854                    & 0.978                    & 0.995                    & 0.883                    & 0.986                    & 0.997                    & 0.790                    & 0.991                    & 0.996                   \\
Efficient-Unet-B1 ($w=1, d=1.1$)                                  & Minimalistic                                        & 0.869                    & 0.980                    & 0.994                    & 0.933                    & 0.973                    & \textbf{0.996}                    & 0.875                    & 0.961                    & 0.987                    & 0.855                    & 0.980                    & 0.995                    & 0.889                    & 0.991                    & \textbf{0.997}                    & 0.792                    & 0.993                    & 0.996                   \\
Efficient-Unet-B2 ($w=1.1, d=1.2$)                                  & Minimalistic                                        & 0.871                    & 0.981                    & 0.994                    & \textbf{0.935}                    & 0.977                    & \textbf{0.996}                    & 0.878                    & 0.962                    & 0.987                    & 0.853                    & 0.977                    & 0.995                    & 0.888                    & 0.992                    & \textbf{0.997}                    & 0.801                    & 0.995                    & \textbf{0.997}                   \\
Efficient-Unet-B3 ($w=1.2, d=1.4$)                                  & Minimalistic                                        & \textbf{0.875}                    & \textbf{0.984}                    & \textbf{0.995}                    & \textbf{0.935}                    & \textbf{0.978}                    & \textbf{0.996}                    & \textbf{0.881}                    & \textbf{0.969}                    & \textbf{0.990}                    & \textbf{0.859}                    & \textbf{0.982}                    & \textbf{0.996}                    & \textbf{0.891}                    & \textbf{0.993}                    & \textbf{0.997}                    & \textbf{0.809}                    & \textbf{0.996}                    & \textbf{0.997}                   \\ \bottomrule 
\bottomrule
\end{tabular}
\end{adjustbox}
\caption{Evaluation results on validation set of Amgad \etal \cite{amgad2019structured} dataset using different automatic and interactive segmentation methods.}
\label{tab:results}
\end{table*}

\subsection{Dataset, preprocessing, and metrics}
The dataset used for model training and validation is from the work of Amgad \etal \cite{amgad2019structured} which contains 151 H\&E stained tissue image regions extracted from WSIs of the same number of triple-negative breast cancer cases acquired from the Cancer Genome Atlas. Tissue regions in this dataset have been annotated with the help of 25 annotators.
In total, there are 20340 regions annotated in 20 categories that are merged into 5 classes: tumour, lymphocytic regions, stroma, necrosis or debris, and others (blood, fat, vessel, etc.) \cite{amgad2019structured}.
In this work, we followed the same region labelling paradigm and used the same image regions and rendered masks in  Amgad \etal \cite{amgad2019structured}
which are split into 82 slides for training and 43 slides for validation.


Following \cite{amgad2019structured}, images are stain normalized using Reinhard's method \cite{reinhard2001color}. The original images are captured at 0.25 micron per pixel (MPP) resolution (which is equal to 40x magnification) with various scanners. However, to keep enough context during the training of our interactive segmentation model, we
extract $512\times512$ patches from image regions and their corresponding masks at 10x magnification (1 MPP resolution). Having guiding signals in the input of the interactive segmentation model, we can use lower resolution images to speed up the region marking and processing time. However, extracted patches were confirmed by a pathologist to show enough contextual and detailed information for tissue region annotation.

For the evaluation of the segmentation performance, we calculate the DICE similarity coefficient, accuracy, and Area Under ROC curve (AUC) of pixel-wise classification. These metrics are reported for each region type separately as well as the overall performance for all tissue types.

\subsection{Validation experiments}

In this section, we report the results of applying the proposed method and other  state-of-the-art (SOTA) interactive or automatic segmentation models on the validation set of the Amgad dataset \cite{amgad2019structured} in \cref{tab:results}, where the first column indicates the segmentation model, the second column designate the guiding signal used for that model training, and the rest six columns report the evaluation metrics on overall performance or each tissue category. Please note that during the tests on the validation set, for all interactive models, similar guiding signals are provided to make fair comparisons.

Results in \cref{tab:results} suggest that interactive segmentation models like NuClick \cite{koohbanani2020nuclick} and the proposed method can outperform SOTA automatic segmentation models like UNet \cite{ronneberger2015unet}, DeepLab v3 \cite{chen2017deeplab}, and the baseline method \cite{amgad2019structured} by a large margin as they are provided with guiding signals in the input. Particularly, our best performing model, \textbf{Efficient-Unet-B3}, achieves overall Dice, accuracy, and AUC of 0.875, 0.984, 0.995, respectively. In comparison to the SOTA automatic segmentation models, our proposed approach performs about 14\% and 11\% better than UNet \cite{ronneberger2015unet} and DeepLab v3 \cite{chen2017deeplab} in terms of overall Dice score, respectively. The same trend can be seen not only for overall accuracy and AUC metrics but also for all the metrics reported for different tissue types in \cref{tab:results}. Note that the lower performance for some region types in comparison to the overall (average) performance can be associated with the higher noise in GT annotations of those regions. Higher noise in GT arises from ambiguity in the boundaries of these regions which makes it hard to separate them from other regions as reported in \cite{amgad2019structured}. Although the original NuClick  \cite{koohbanani2020nuclick} performs better than all other automatic segmentation models (overall Dice score 0.773), it still shows lower performance metrics than the proposed method i.e., Efficient-Unet-B3 segmenting 10\% better in terms of Dice score. Nevertheless, it can be seen that when we train the NuClick model with the proposed minimalistic signals (\cref{sec:skeleton}), overall Dice scores rises to 0.835, which shows the effectiveness of the proposed minimalistic guiding signal generation.

The four last rows of \cref{tab:results} are reporting the effect of increasing the scale of the proposed Efficient-UNet model architecture. In each row, number of features maps $F$ and repetition of MIRSE blocks $R$ (see \cref{tab:net}) are multiplied by the introduced $w$ and $d$ factors.
Scaling up the network from Efficient-Unet-B0 variant to the B3 variant does elevate the performance level of interactive segmentation, however, this raise is not significant. Specifically, the largest model (Efficient-Unet-B3) performs only 0.8\% better than the baseline model (Efficient-Unet-B0) in overall Dice.


\subsection{Experiment with human-drawn guiding signals}

\begin{table}[]
\begin{adjustbox}{max width=\columnwidth}
\begin{tabular}{@{}lllll@{}}
\toprule
\toprule
\multicolumn{1}{c}{\multirow{2}{*}{Model}}   & \multicolumn{1}{c}{\multirow{2}{*}{Guiding Signal}} & \multicolumn{3}{c}{Overall}                                                   \\ \cmidrule(l){3-5} 
\multicolumn{1}{c}{}                         & \multicolumn{1}{c}{}                                & \multicolumn{1}{c}{DICE} & \multicolumn{1}{c}{Acc.} & \multicolumn{1}{c}{AUC} \\ \midrule
\multicolumn{1}{l}{Efficient-Unet-B0}       & \multicolumn{1}{l}{Minimalistic}                       & 0.874                    & 0.982                    & 0.994                   \\
\multicolumn{1}{l}{Efficient-Unet-B0}       & \multicolumn{1}{l}{Human \#1}                          & 0.887                    & 0.986                    & 0.995                   \\
\multicolumn{1}{l}{Efficient-Unet-B0}       & \multicolumn{1}{l}{Human \#2}                          & 0.870                    & 0.981                    & 0.994                   \\ \bottomrule
\bottomrule
\end{tabular}
\end{adjustbox}
\caption{Comparison of proposed model performance on a subset of 25 validation images when used with auto-generated minimalistic guiding signals versus human-provided guiding signals.}
\label{tab:results_human}
\end{table}

To test the robustness of the proposed method against the variations in human-provided guiding signals, we presented a subset of validation set (25 images) to 2 non-expert annotators and asked them to draw guiding signals on images. To make the process consistent between the two annotators, for each image and each region type we overlaid the ground truth annotation on images and asked the annotators to draw their squiggle markers (guiding signals) inside the annotated region. By doing this, we are able to compare the performance of the same model using minimalistic (auto-generated) signals (\cref{sec:skeleton}) versus using human-drawn guiding signals as reported in \cref{tab:results_human}.

The trained Efficient-Unet-B0 performs accurately whether the input guiding signals are automatically generated or provided by a human annotator.
In \cref{tab:results}, the model provided with guiding signals drawn by annotator ``Human \#2''  interestingly performs better than the same model with auto-generated signals, achieving an overall Dice score of 0.887 over 0.874. Qualitative results for three different samples in \cref{fig:results} (where each region type has been coloured differently and guiding signals for tissue region are depicted with dark blue lines) also show that the trained models can handle variations in the guiding signals and segment tissue regions quite perfectly even if provided with overly simplified guiding signals.

\section{Discussion}
\label{sec:discussion}

\begin{figure*}[th!]
\begin{center}
\includegraphics[width=0.99\linewidth]{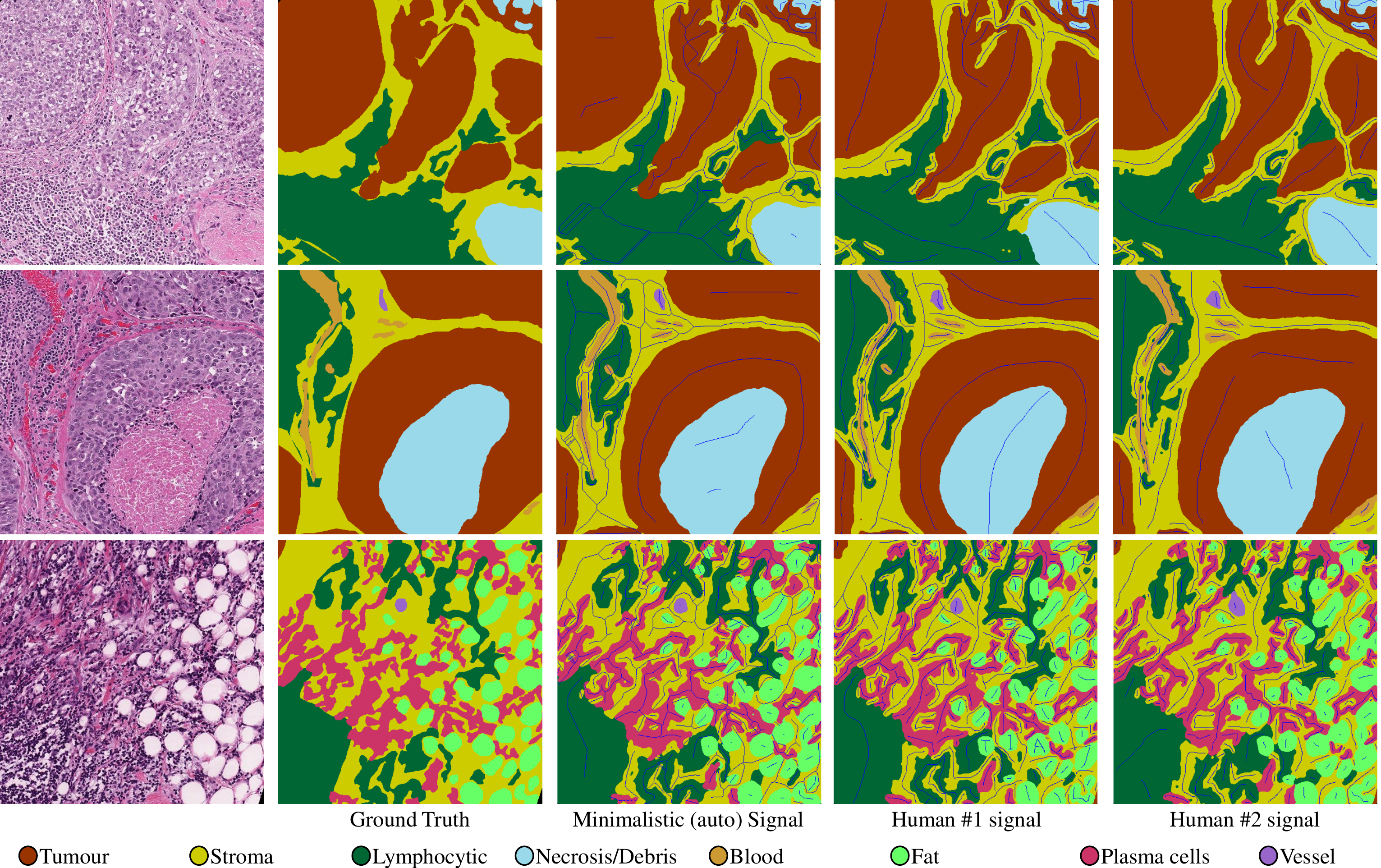}
\end{center}
   \caption{Results of applying proposed baseline interactive segmentation model, \textbf{Efficient-UNet-B0}, on three samples using minimalistic guiding signals (described in \cref{sec:skeleton}), and guiding signals provided by two human annotators.}
\label{fig:results}
\end{figure*}


Evaluation results in \cref{tab:results} show that our proposed segmentation models, Efficient-UNet B0-B3, perform considerably better than SOTA interactive segmentation model, NuClick \cite{koohbanani2020nuclick}, using the same minimalistic guiding signal (\cref{sec:skeleton}) and increase the overall Dice metric by 3\%--4\% as the network scale grows. One may argue that despite the increase of network scale, its performance does not elevate considerably i.e., less than 1\% increase in all performance metrics throughout all tissue types when increasing the Efficient-UNet scale from B0 to B3. With the growth of network scale, our models capacity for learning feature representations also expands \cite{lecun2015deep}, however, bigger networks usually need more training data to be trained optimally whereas we only used a limited number of training images in this work. Therefore, we believe that one of the main reasons for not having a considerable performance boost while increasing network scale is the lack of training data. Nevertheless, we kept the design of the network as scaleable as possible so it can be used in future experiments or other applications with more data. 

The efficiency of the proposed segmentation network is not only related to its accuracy but also its speed. The baseline network Efficient-UNet-B0 is twice faster as compared to NuClick while gaining 3\% higher overall Dice metric. Particularly, our network can generate the annotation map for each guiding signal in about 0.1 seconds (using an Nvidia Tesla V100 GPU). This means that the proposed network fits in an interactive annotation setup where it can respond to the annotator's markers in real-time and make the interactive process more user friendly. 

The positive effect of introducing proposed minimalistic signal generation techniques during the training phase is visible in validation results of \cref{tab:results} and \cref{tab:results_human}. In \cref{tab:results}, one can see that by only using the proposed minimalistic guiding signal generation method, overall NuClick performance (Dice metric) increases by more than 6\% that is more prominent than the effect of improving network architecture discussed in the previous paragraphs (3\% improvement when using Efficient-Unet instead of NuClick). Results in \cref{tab:results_human} also testify to the effectiveness of the proposed method for guiding signal generation, where a given model performs at the same level or even better when it is provided human-drawn guiding signals instead of GT-based extracted ones. This shows that the proposed techniques in \cref{sec:skeleton} can generate guiding signals that resemble human-drawn markers and capture the variations they might show. This claim is supported by the qualitative results illustrated in \cref{fig:results}, where even overly-simplified squiggles as guiding signals can lead to high quality segmentation outputs. This also implies that the proposed interactive segmentation model is insensitive to variations in the user input. This experiment expresses the adaptiveness of the proposed interactive segmentation method, where the more carefully and detailed guiding signals are drawn, the more accurate the segmentation output will be. Having near perfect initial annotations by only providing a simple guiding signal in the beginning will save a considerable amount of annotator time during the annotation review and correction \cite{wahab2021semantic}.

Although we have trained and tested our interactive segmentation models on image-level information, it is only a matter of implementation to extend this idea to work on WSIs. Coupled with WSI viewer and annotation tools like ASAP \cite{litjens2017asap}, the proposed interactive method can speed up the annotation curation process considerably. Annotators only need to view the WSI at low-level magnification and draw squiggles with different indices (colours) inside different tissue types, then we only need to extract patches based on the drawn markers and feed them into the proposed interactive segmentation model to get the initial segmentation map for each path. Results of the patches are then stitched and overlaid on the WSI in the form of polygons so they can be edited by the annotators.  

\section{Conclusions and future directions}
\label{sec:conclusion}
In this work, we proposed an interactive semantic segmentation model for robust tissue region annotation in a semi-automated manner. Our approach follows the work of \cite{koohbanani2020nuclick} in which for segmentation of each object in the image, two guiding signal maps (inclusion and exclusion maps) are concatenated to the image in the input of the network. Here, we improved the network architecture by introducing Efficient-UNet which is 2 times faster and 4\% more accurate in terms of overall Dice metric than the original NuClick. We have also proposed four novel techniques to automatically extract minimalistic and human-drawn-like guiding signals from GT masks, so they can be used during the training of the model. We showed that using the proposed minimalistic guiding signals makes the models robust against variations in the user input and can improve the overall Dice metric by another 6\%. We believe this AI tool can be embedded in a whole slide image viewer/annotation software efficiently to effectively accelerate large-scale region annotation acquisition projects.

There are several ideas that can be pursued to extend this work. Testing with the generalizability of the proposed method to new unseen datasets (e.g. samples from other organs) or checking the adaptability of the trained model to new domains (like IHC stained samples) can be of interest as an extension of this work. In that case, we would like to investigate the effect of incorporating unsupervised pre-training methods, such as SelfPath \cite{koohbanani2021self}, on the performance of the proposed interactive method for tissue region segmentation in new domains. Furthermore, it is possible to add a classification head to the proposed network architecture to enable proposing a category label for each region which may also improve the segmentation quality synergistically.

{\small
\bibliographystyle{unsrt}
\bibliography{egbib}
}

\end{document}